\documentclass{article}
\usepackage{ijcai21}

\usepackage{times}
\usepackage{soul}
\usepackage{url}
\usepackage[hidelinks]{hyperref}
\usepackage[utf8]{inputenc}
\usepackage[small]{caption}
\usepackage{graphicx}
\usepackage{amsmath, amssymb}
\usepackage{amsthm, bm}
\usepackage{booktabs}
\usepackage{algorithm}
\usepackage{algorithmic}
\usepackage{xcolor,color}
\usepackage{subcaption}

\title{Simulation of Electron-Proton Scattering Events by a Feature-Augmented \\ and Transformed Generative Adversarial Network (FAT-GAN)}

\author{
Yasir~Alanazi$^1$\and
N.~Sato$^2$\and
Tianbo~Liu$^2$\and
W.~Melnitchouk$^2$\and
Pawel Ambrozewicz$^2$\and\\
Florian Hauenstein$^3$\and
Michelle~P.~Kuchera$^4$\and
Evan~Pritchard$^4$\and
Michael~Robertson$^5$\and\\
Ryan~Strauss$^5$\and
Luisa~Velasco$^6$\And
Yaohang~Li$^1$
\affiliations
$^1$Department of Computer Science, Old Dominion University, Norfolk, Virginia 23529\\
$^2$Jefferson Lab, Newport News, Virginia 23606 \\
$^3$Department of Physics, Old Dominion University, Norfolk, Virginia 23529\\
$^4$Department of Physics, Davidson College, Davidson, North Carolina 28035\\
$^5$Department of Mathematics and Computer Science, Davidson College, Davidson, North Carolina 28035\\
$^6$Department of Physics, University of Dallas, Irving, Texas 75062
\emails
yalan001@odu.edu,
\{nsato, liutb, wmelnitc, pawel, hauenst\}@jlab.org,
\{mikuchera, evpritchard, mirobertson, rystrauss\}@davidson.edu,
lvelasco@udallas.edu,
yaohang@cs.odu.edu
}

\begin{document}

\maketitle

\begin{abstract}
We apply generative adversarial network (GAN) technology to build an event generator that simulates particle production in electron-proton scattering that is free of theoretical assumptions about underlying particle dynamics.
The difficulty of efficiently training a GAN event simulator lies in learning the complicated patterns of the distributions of the particles physical properties.
We develop a GAN that selects a set of transformed features from particle momenta that can be generated easily by the generator, and uses these to produce a set of augmented features that improve the sensitivity of the discriminator.
The new Feature-Augmented and Transformed GAN (FAT-GAN) is able to faithfully reproduce the distribution of final state electron momenta in inclusive electron scattering, without the need for input derived from domain-based theoretical assumptions.
The developed technology can play a significant role in boosting the science of existing and future accelerator facilities, such as the Electron-Ion Collider.
\end{abstract}

\section{Introduction}

High-energy scattering reactions typically produce collections of particles in the final state whose momentum distributions are governed by fundamental femtometer-scale physics. 
One of the major goals of existing lepton-hadron scattering facilities, such as COMPASS at CERN and Jefferson Lab, as well as the future Electron-Ion Collider, is to determine the three-dimensional distributions of the hadrons' elementary quark and gluon constituent from measurements of particle production cross sections. 
Unfortunately, since quarks and gluons are not directly detectable experimentally, their properties must be inferred indirectly from the observed particle spectra within the theoretical framework of factorization in Quantum Chromodynamics (QCD)~\cite{Collins:1989gx}.

Since the early 1970s, Monte Carlo event generators (MCEGs) have played a vital role in facilitating studies of high-energy scattering reactions. 
From the experimental perspective, MCEGs are crucial for understanding the complex arrays of detectors that measure the energies and momenta of final state particles. 
The construction of existing MCEGs, such as Pythia~\cite{Sjostrand:2007gs}, Herwig~\cite{Bahr:2008pv} or Sherpa~\cite{Gleisberg:2008ta}, has been driven by a combination of high-precision data from previous experiments and inputs from theory.
The latter have involved a mix of perturbative QCD methods, describing the dynamics of quarks and gluons at short distances, and phenomenological models that map the transition from quark and gluon degrees of freedom to the observable hadrons (``hadronization'').
An MCEG can in principle be viewed as a form of a ``data compatification tool'', encapsulating enormous amounts of data collected from multiple experiments, which can be regenerated from the MCEG itself.
On the other hand, the reliance of existing MCEGs on theoretical assumptions of factorization and on hadronization models limits their ability to capture the full range of possible correlations between produced particles' momenta and spins.

In this work we suggest a new strategy for constructing MCEGs using modern machine learning methods involving GANs~\cite{Goodfellow:2014} that can learn to generate particles in specific reactions, such as electron-proton scattering, without recourse to theoretical assumptions about femtometer-scale physics. 
A particular feature of GANs is their ability to generate synthetic data, such as images, by learning from real samples without knowledge of the underlying laws of the original system.
The aim is to implicitly learn the primordial distributions over data which are difficult to model with an explicit parametrization for the underlying law. 
We propose to explore this aspect by treating the events characterized by final state particle momenta in high-energy reactions as the ``images''.

Typically, a GAN model is composed of a generator and a discriminator. 
The generator transforms random white noise through a deep neural network to produce candidate samples from the target distribution, while the discriminator learns through another deep neural network to differentiate the true samples from those produced by the generator.
The GAN evolves as the generator and discriminator compete adversarially, alternatively updating their parameters during the training process. 
Eventually, the GAN is able to approximate the underlying cumulative distribution function (CDF) and inverse CDF transformation, and thus sample the target distribution given sufficiently large neural networks, sample size, and long enough computation time.

Although GANs have demonstrated impressive results in generating near-realistic images~\cite{karras2018stylebased}, music~\cite{mogren2016crnngan}, and videos~\cite{clark2019adversarial}, training a successful GAN model is known to be notoriously difficult.
Many GAN models suffer from major problems including 
mode collapse,
non-convergence, 
model parameter oscillation, 
destabilization,
vanishing gradient, 
and overfitting due to unbalanced generator/discriminator combinations. 
Approaches and techniques to address these general problems have been proposed and discussed in a number of publications \cite{Salimans2016ImprovedTF,Arora2017DoGA,Arjovsky2017TowardsPM}.

Using GANs to simulate events from particle reactions poses additional challenges for machine learning.
Unlike many GAN applications, such as generating realistic and sharp looking images, where the distribution agreement between the GAN-generated samples and the true ones is often not strictly enforced, GANs for generating physical events are required to model the distributions of event features and their correlations sufficiently precisely for the nature of particle reactions to be faithfully replicated.
Furthermore, events generated by GANs should not violate basic physical laws, such as baryon number, charge and momentum conservation.
In order for the GAN event generator to work, a careful choice of architecture, representations, features, parameter initialization, and selection of hyper-parameters is required.

In this paper we describe the development of a GAN event generator to simulate particles in inclusive electron-proton scattering.
As a proof of concept, we restrict ourselves to building a GAN that only learns how to generate specific kinds of particles in the final state, while ignoring other particles.
We shall refer to such a generator as an ``inclusive'' generator, to distinguish it from an ``exclusive'' generator that generates the full spectrum of particles in the final sate. 
For testing and validation purposes we first use the existing theory-based Pythia8 MCEG \cite{Sjostrand:2007gs} to generate the training samples to train GAN. Then we train and validate GAN on data from real-life electron scattering experiment.

Our analysis shows that the difficulty of successfully training a GAN event generator lies in the complicated patterns of distributions of the physical properties of the generated particles.
To improve the GAN training, we develop a Feature-Augmented and Transformed GAN (FAT-GAN), using a set of transformed features from the particle physical properties as the generated features of the generator. 
These transformed features represent the underlying degrees of freedom of the particles, which can simplify the learning of the generator, while avoiding generation of unphysical events.
A set of augmented features is further derived from the generated features to improve the sensitivity of the discriminator. 
We also compare the efficiency of using Cartesian coordinates versus spherical coordinates to represent generated features in the FAT-GAN.
With well-selected coordinate systems and features, our GAN model is able to regenerate particle momenta so that it mimics all the relevant momentum correlations as observed in the the original MCEG.

\section{Related work}

In the literature, GANs have been used in a variety of applications at the Large Hadron Collider (LHC), such as simulating energy deposits of final state particles \cite{Paganini_2018,CaloGAN_2018} and jets \cite{de_Oliveira_2017,Musella_2018}, accelerating importance sampling Monte Carlo integration \cite{bendavid2017efficient}, simulating data collections of beam studies \cite{Erdmann_2019}, and reconstructing cosmic ray-induced air showers \cite{Erdmann_2018}. 
These efforts have focused on image representations of the aggregated experimental data as opposed to individual event-based samples. 
GANs have also been used to invert the detector effects such as in \cite{datta2018unfolding} and \cite{Bellagente_2020}.
A crucial question about whether the events generated by GANs can add statistical precision beyond the training samples was investigated in \cite{butter2020ganplifying}.

Recently there have been several attempts to investigate the possibility of directly training GANs at the event level in proton-proton collisions, such as those at the LHC.
%
\cite{hashemi2019lhc} applied a GAN to produce muon four-momenta in $Z \to \mu^+\mu^-$ events generated by Pythia.
Although agreeing well in the reduced dataset, their GAN model fails to reproduce certain feature distributions. 
%
\cite{otten2019event} reported a less satisfactory performance when a GAN was based on fully-connected deep networks in the study of the two-body decay processes $e^+ e^- \to Z \to l^+ l^-$ and $pp \to t\bar t$.
%
\cite{butter2019GAN} applied a GAN to simulate the $2 \to 6$ particle production process $pp \to t \bar{t} \to (bq\bar{q}')(\bar{b}\bar{q}q')$, incorporating maximum mean discrepancy (MMD)~\cite{MMD_2007} to resolve sharp local features. 
%
\cite{DiSipio:2019imz} implemented a dijet-GAN based on convolutional neural networks (CNNs) to simulate the production of pairs of jets at the LHC.

Although each of the above GANs was designed to simulate events in different reactions, they all face similar challenges of learning the unique feature distributions of the events, which often exhibit sharp edges, spikes, multiple peaks, and large variations. 
Building a generator capable of generating these distributions and a discriminator sensitive to their features are the keys to the success of GAN-based MCEGs.

\section{Methods}

\subsection{Data descriptions}

We use GANs to mimic two samples of inclusive electron-proton scattering events: the first is generated from the Pythia MCEG \cite{Sjostrand:2007gs} at a center-of-mass energy of 100~GeV, and the second is experimental data from CLAS at Jefferson Lab with beam energy of 5.5~GeV.
In our initial analysis, we train the GAN only on the scattered electron momenta.
Events are represented as an array of the electron four-momenta 
    $p_\mu = (E; \bm{p})$, 
where in Cartesian coordinates the three-momentum is given by 
    $\bm{p} = (p_x, p_y, p_z)$,
and the energy is given by
    $E = \sqrt{p_x^2 + p_y^2 + p_z^2 + m^2}$,
where $m$ is the electron mass.
Throughout this paper we will work in units of GeV for all momentum and energy variables.

\subsection{GAN architecture}
\label{GANarchitecture}

The generator produces both generated features and augmented features. 
The three generated features describe the three degrees of freedom of the scattered electron (components of the momentum $\bm{p}$), while the augmented features, such as energy $E$, transverse momentum 
    $p_T = \sqrt{p_x^2+p_y^2}$,
and the longitudinal to transverse momentum ratio $p_z/p_T$, can be calculated from the generated features to represent other physical characteristics of the particle. 
The augmented features are used to improve the sensitivity of the discriminator.

The input to the generator ``$G$'' is a $100$-dimensional white noise array centered at $0$ with unit standard deviation.
The generator network consists of $5$ hidden dense layers, each with $512$ neurons, activated by a leaky Rectified Linear Unit (ReLU) function.
The last hidden layer is fully connected to a three-neuron output, activated by a linear function representing the generated features. 
A customized Lambda layer is then incorporated to calculate the augmented features from the generated features. 
Inspired by the idea of importance sampling~\cite{ji2016}, the augmented features are carefully selected to improve the sensitivity of the discriminator in distinguishing the GAN-generated events from the Pythia input.
These augmented features, together with the generated features from the generator, are concatenated and fed to the discriminator as input.

As for the generator, the neural network in the discriminator ``$D$'' also consists of $5$ hidden dense layers, each with $512$ neurons, activated by a leaky ReLU function.
To avoid overfitting in classification, a $10\%$ dropout rate is applied to each hidden layer. 
The last hidden layer is fully connected to a single-neuron output, activated by a sigmoid function, where ``$1$'' indicates a true event and ``$0$'' is a fake event.
The overall architecture of the inclusive GAN event generator is illustrated in Fig.~\ref{fig:gan}.

\begin{center}
\includegraphics[width=0.85\linewidth]{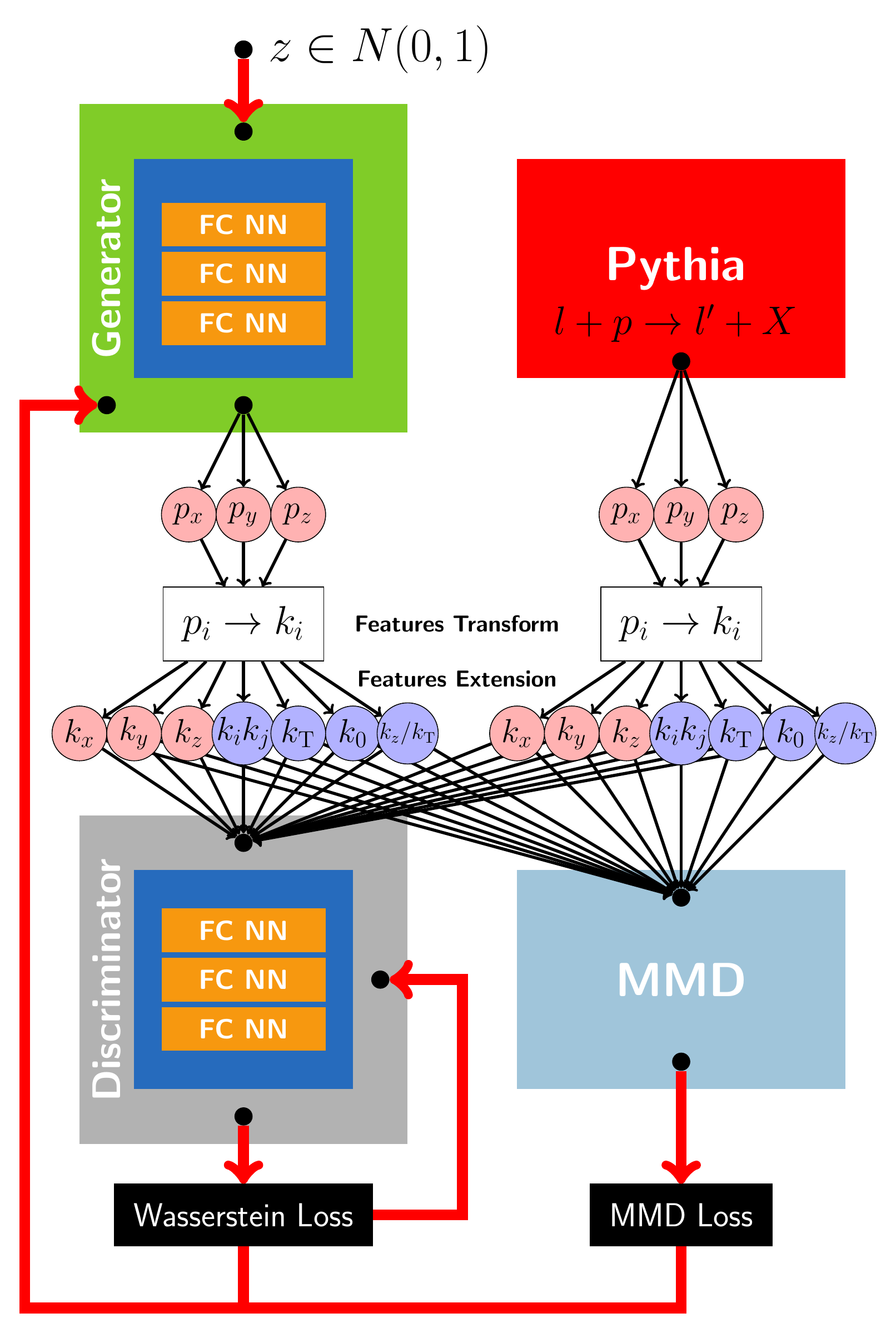}
\captionof{figure}
{Architecture of the inclusive FAT-GAN event generator (see text for details).} 
\label{fig:gan}
\end{center}

\subsection{Loss functions} 

The discriminator $D$ is trained to give $D(\bm{p}) = 1$ for each sample $\bm{p}$ generated by Pythia, and $D(\widetilde{\bm{p}}) = 0$ for each sample $\widetilde{\bm{p}}$ produced by the generator.
The discriminator is optimized against the Wasserstein loss with gradient penalty \cite{WGAN_2017} to improve training stability and reduce the likeliness of mode collapse. 
The loss function $L_D$ of the discriminator is defined as
\begin{equation}
\begin{aligned}
L_D & = \big( \mathbb{E}[D(\widetilde{\bm{p}}))] - \mathbb{E}[D(\bm{p})] \big) \\
    & + \lambda\, \mathbb{E}_{\bm{\widehat p} \sim P_{\bm{\widehat p}} }
                \big[(\|\nabla_{\!\bm{\widehat p}}D(\bm{\widehat p})\|_2-1)^2\big],
\label{eq.L_D}
\end{aligned}
\end{equation}
where $\mathbb{E}$ denotes the expectation value.
The first term in Eq.~(\ref{eq.L_D}) measures the Wasserstein distance \cite{arjovsky2017wasserstein}. 
The second term is the gradient penalty, where $\bm{\widehat p}$ is a random sample from $P_{\bm{\widehat p}}$, defined by a uniform distribution along the straight lines between pairs of samples from the true Pythia event data and the generator's output. 
The coefficient $\lambda$ is a harmonic parameter to balance the Wasserstein distance and the gradient penalty.

\begin{figure*}[ht]
\centering
\includegraphics[width=0.70\textwidth]{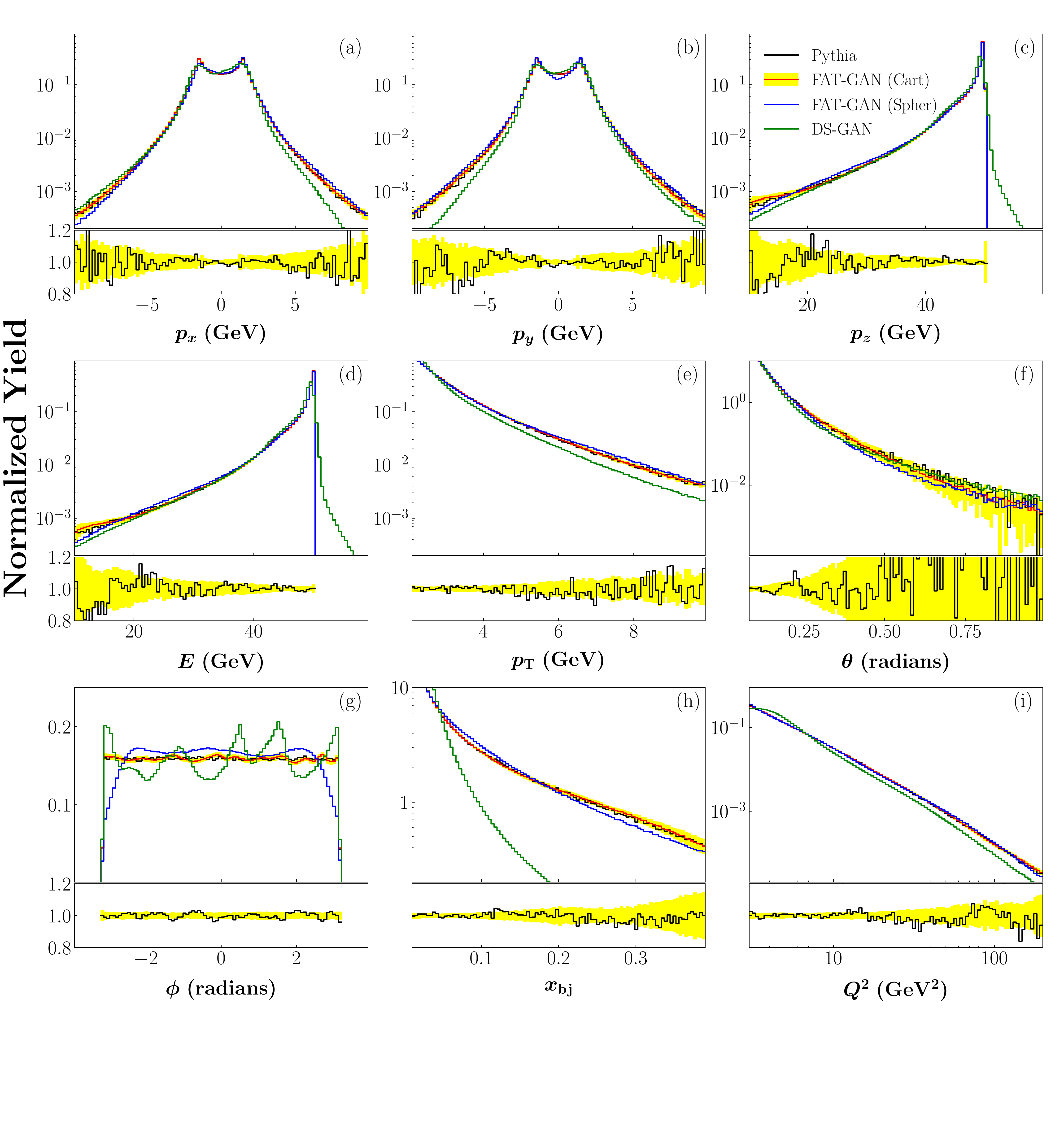}
\vspace*{-1.5cm}
\caption{Distributions of physical properties of the scattered electron, $p_x$, $p_y$, $p_z$, $E$, $p_T$, $\theta$, $\phi$, $x_{\rm bj}$ and $Q^2$ (see text), generated by Pythia (black lines), FAT-GAN (Cart) (red lines and yellow bands), FAT-GAN (Spher) (blue lines), and DS-GAN (green lines). The ratio of the FAT-GAN (Cart) to Pythia yields is shown at the bottom of each panel.}
\label{fig:GANresults}
\end{figure*}

To ensure that the distributions of the event features created by the generator match with those of Pythia, we incorporate a two-sample test based on kernel MMD in our inclusive GAN event generator. 
To compare two distributions, the MMD employs a kernel-based statistical test method to determine if the two samples are drawn from different distributions. 
As a result, the loss function $L_G$ of the generator $G$ includes a Wasserstein distance term from the discriminator $D$ and an MMD term \cite{li2017mmd},
\begin{equation}
\begin{aligned}
L_G = -\mathbb{E}[D(\widetilde{\bm{p}}))] + \eta\, \textrm{MMD}^2(\bm{p},\widetilde{\bm{p}}),
\end{aligned}
\end{equation}
where $\eta$ is the balancing hyperparameter. 
The MMD term is defined as
\begin{equation}
\begin{aligned}
\textrm{MMD}^2(\bm{p},\widetilde{\bm{p}})
& = \mathbb{E}_{\bm{p}_a, \bm{p}_{a'} \sim P_{\bm{p}}}[k(\bm{p}_a, \bm{p}_{a'})] \\
& +\, \mathbb{E}_{\bm{p}_b, \bm{p}_{b'} \sim P_{\widetilde{\bm{p}}}}[k(\bm{p}_b, \bm{p}_{b'})] \\
& - 2\, \mathbb{E}_{\bm{p}_a \sim P_{\bm{p}}, \bm{p}_b \sim P_{\widetilde{\bm{p}}}}[k(\bm{p}_a, \bm{p}_b)],
\end{aligned}
\end{equation}
where $k(\bm{p}_a, \bm{p}_b)$ is a positive definite kernel function.
Here we select a Gaussian kernel such that 
$k(\bm{p}_a, \bm{p}_b) = \exp[-(\bm{p}_a-\bm{p}_b)^2 / 2\sigma^2]$, 
where $\sigma$ is the hyperparameter determining the MMD resolution, tuned to the same order of magnitude as the width of the event features.

The combined network is trained adversarially for $200,000$ epochs. 
In each epoch, the combined network passes through the complete training dataset once.
A large batch of $10,000$ events is employed to ensure that there are sufficient samples to calculate a stable MMD value in each batch. 
Each batch contains random examples from the Pythia event dataset. 
The optimizer is Adam \cite{kingma2014adam} with a $10^{-4}$ learning rate, $\beta_1 = 0.5$, and $\beta_2 = 0.9$.
To balance the generator and discriminator training, the training ratio is set to~$5$.

\subsection{Feature representations}

The physical observables characterizing the scattered electron properties are illustrated in Fig.~\ref{fig:GANresults}, including the momentum and energy components of the electron four-vector $p_\mu$ [Fig.~\ref{fig:GANresults}(a)--(e)], scattering angles [Fig.~\ref{fig:GANresults}(f)--(g)], and derived quantities $x_{\rm bj}$ and $Q^2$ (see Sec.~\ref{ssec:inter-correlations} below).
The energy and momentum distributions generated by Pythia exhibit rather large variations, with the ratio between the most populated regions and those with rare events reaching up to $\sim 10^4$.

More seriously, a sharp edge in the $E$ distribution arises from energy conservation, which restricts $E$ to be less than the incident beam energy, $E_{\rm b}$.
This sharp edge is very difficult for the inclusive GAN to learn, as unphysical events can be generated with $E > E_{\rm b}$, which the discriminator is not sensitive enough to differentiate from the eligible physical events, particularly when $E_{\rm b} - E$ is small. 
The sharp edge in the $E$ distribution also leads to sharp edges in the $p_z$ and $\theta$ distributions.
The difficulty of learning sharp edge distributions has also been reported in \cite{hashemi2019lhc}.

To address the problem of learning sharp edge distributions, we transform the momentum properties to specific generated features that allow their distributions to be generated more easily, while avoiding production of unphysical particles.
For the $p_z$ distribution, instead of directly using $p_z$ as a generated feature, we use the transformed variable
    ${\cal T}(p_z) = \log[(E_{\rm b} - p_z)/(1~{\rm GeV})]$
in the generator.
The original distribution with a sharp edge is now converted to a distribution that is more like a Gaussian, with significantly reduced variation, as illustrated in Fig.~\ref{fig:lnpz}.

\begin{center}
\includegraphics[width=0.65\linewidth]{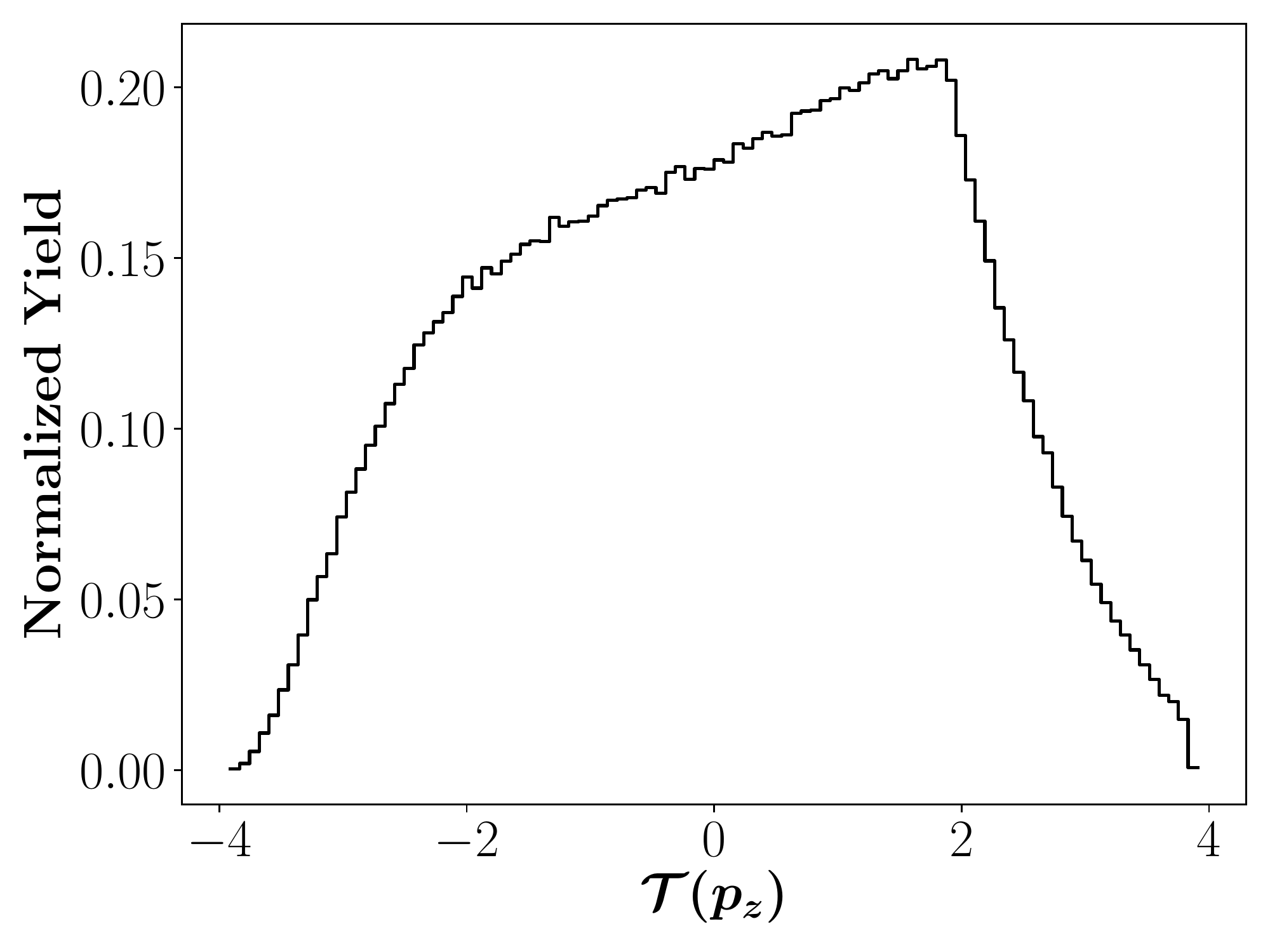}
\vspace*{-0.2cm}
\captionof{figure}{Distribution of the transformed feature ${\cal T}(p_z)$.}
\label{fig:lnpz}
\end{center}

Although ${\cal T}(p_z)$ does not have actual physical meaning, the transformation ensures that the GAN will not generate events with unphysical $p_z$ values.
As well as making it easier for the generator to produce, the transformed distribution ${\cal T}(p_z)$ improves the sensitivity of the discriminator as a classifier.
As a result, the generator learns to generate $(p_x, p_y, {\cal T}(p_z))$, and $p_z = \exp[({E_{\rm b}-{\cal T}(p_z)})/(1~{\rm GeV})]$ is later calculated by the customized Lambda layer as one of the augmented features.

In addition to $p_z$, the variables $p_T$, $E$, and $p_z/p_T$ are also calculated in the customized Lambda layer as augmented features in order to improve the sensitivity of the discriminator. 
The generated features and augmented features are concatenated as the input to the discriminator.

\section{Results}

In this section we compare the event feature distributions from the FAT-GAN and the Direct-Simulation GAN (``DS-GAN'') that directly simulates the momenta ($p_x$, $p_y$, $p_z$).
We also compare the efficiency of the FAT-GAN implementation in Cartesian coordinates (``FAT-GAN (Cart)'') with the FAT-GAN in spherical coordinates (``FAT-GAN (Spher)'').
The latter is an alternative representation to describe a particle in terms of the variables ($E$, $\theta$, $\phi$), where 
    $\theta = \arctan(p_z/p_T)$
is the polar angle between $p_z$ and the transverse plane, and
    $\phi = \arctan(p_y/p_x)$
is the azimuthal angle.
The FAT-GAN (Spher) generates (${\cal T}(E)$, ${\cal T}(\theta)$, $\phi$) in spherical coordinates as the generated features, and then $p_x$, $p_y$, $p_z$, $p_T$, $E$, and $p_z/p_T$ as the augmented features. 
The features ${\cal T}(E)$ and ${\cal T}(\theta)$ are converted from $E$ and $\theta$ using a logarithmic transformation similar to that applied to $p_z$ to remove the sharp edges in the distribution.

\subsection{Feature distributions}

The event feature distributions from the DS-GAN, FAT-GAN (Cart), and FAT-GAN (Spher) are compared in Fig.~\ref{fig:GANresults} with those generated from Pythia.
In general the DS-GAN results do not match as well with Pythia compared to FAT-GAN (Cart) and FAT-GAN (Spher). 
Moreover, the tails beyond the sharp edges in Fig.~\ref{fig:GANresults}(c) and (d) indicate that some unphysical events are generated with $E > 50$~GeV.
The FAT-GAN (Cart) yields match better with Pythia than do the FAT-GAN (Spher) yields, particularly for the $\phi$ distribution. 
The four momentum components ($E$; $p_x$, $p_y$, $p_z$), as well as $p_T$, $\theta$ and $\phi$, are also well reproduced relative to Pythia, and have a minimal number of unphysical events.

Compared to the reaction events simulated in \cite{butter2019GAN} and \cite{DiSipio:2019imz}, where the typical ratio between the peak and tail events is up to $10$, the ratio in our scattering electron features can be up to $10^4$. 
Nevertheless, even for the rare events that are $10^{-3}$ of the number of the peak events, the FAT-GAN (Cart) agrees well with Pythia in their distributions, including the symmetry of $\phi$ shown in Fig.~\ref{fig:GANresults}.

\subsection{Inter-correlations between event features}
\label{ssec:inter-correlations}

In addition to the electron momentum and energy, we examine two additional physical quantities that are typically used to characterize electron scattering, namely the squared four-momentum of the exchanged virtual photon $Q^2$ and the Bjorken scaling variable $x_{\rm bj}$, neither of which are explicitly generated as features in the FAT-GAN.
In terms of the beam and scattered electron momenta and energies, the photon virtuality can be written as
%
$-Q^2 \equiv q \cdot q = q_0^2 - \bm{q}^2$,
%
where
    $q_0 = E_{\rm b} - E$
and
    $\bm{q} = (-p_x, -p_y, \sqrt{E_{\rm b}^2-m^2}-p_z)$
are the energy and three-momentum transfer, respectively.
The Bjorken variable is defined by the dimensionless ratio
\begin{equation}
\begin{aligned}
x_{\rm bj} = \frac{Q^2}{2 P \cdot q},
\end{aligned}
\end{equation}
and kinematically ranges from $0$ to $1$.
%
%
As shown in Fig.~\ref{fig:GANresults}(h) and (i), the $x_{\rm bj}$ and $Q^2$ distributions from the DS-GAN deviate significantly from those generated by Pythia. In contrast, the FAT-GAN (Cart) yields match better than those from the DS-GAN and the FAT-GAN (Spher).

\begin{center}
\includegraphics[width=0.9\linewidth]{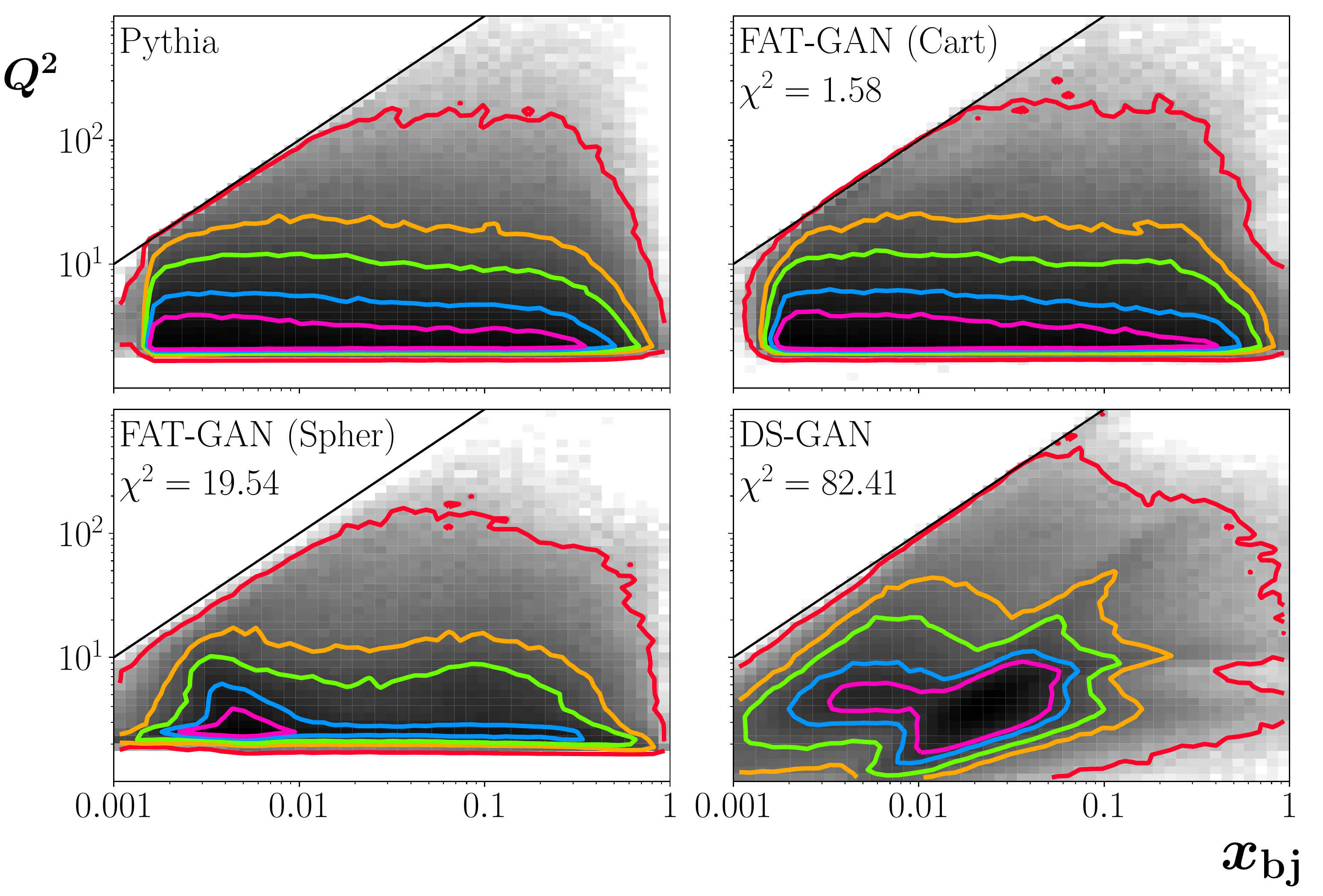}
\vspace*{-0.2cm}
\captionof{figure}{Joint distributions of $Q^2$ (in GeV$^2$) and $x_{\rm bj}$ for FAT-GAN (Cart), FAT-GAN (Spher) and DS-GAN compared to Pythia. The $\chi^2$ values per bin are indicated in the GAN-generated panels.} 
\label{fig:Q2vsxbj}
\end{center}

The $Q^2$--$x_{\rm bj}$ joint distributions, shown in Fig.~\ref{fig:Q2vsxbj}, indicate that the FAT-GAN (Cart) gives a good match with the Pythia $Q^2$-$x_{\rm bj}$ joint distribution, as indicated by the contour lines, with a $\chi^2$ value per bin of $1.58$.
In contrast, somewhat worse agreement ($\chi^2 = 19.54$) is observed for the FAT-GAN (Spher), and very poor agreement ($\chi^2 = 82.41$) for the DS-GAN. 
This indicates that the FAT-GAN model in Cartesian coordinates not only learns the four-momentum vector accurately, but also their inter-correlations.

\subsection{Comparison of coordinate representations}

The Cartesian representation and the spherical representation of a particle are equivalent physically. 
However, the Cartesian and spherical representations exhibit different learning efficiencies in the FAT-GAN.
As shown in Figs.~\ref{fig:GANresults} and \ref{fig:Q2vsxbj}, the FAT-GAN (Cart) demonstrates better agreement in the distributions of physical properties than the FAT-GAN (Spher).
In Fig.~\ref{fig:chi2} the convergence of the FAT-GAN (Cart), FAT-GAN (Spher) and DS-GAN is compared, as measured by the $\chi^2$ values for the $x_{\rm bj}$ distributions of Pythia and GAN-generated events along training epochs.
Although both FAT-GANs yield better convergence than the DS-GAN, the FAT-GAN (Cart) demonstrates better efficiency and generally lower $\chi^2$ values that are close to $1$ than the FAT-GAN (Spher).

\begin{center}
  \includegraphics[width=0.8\linewidth]{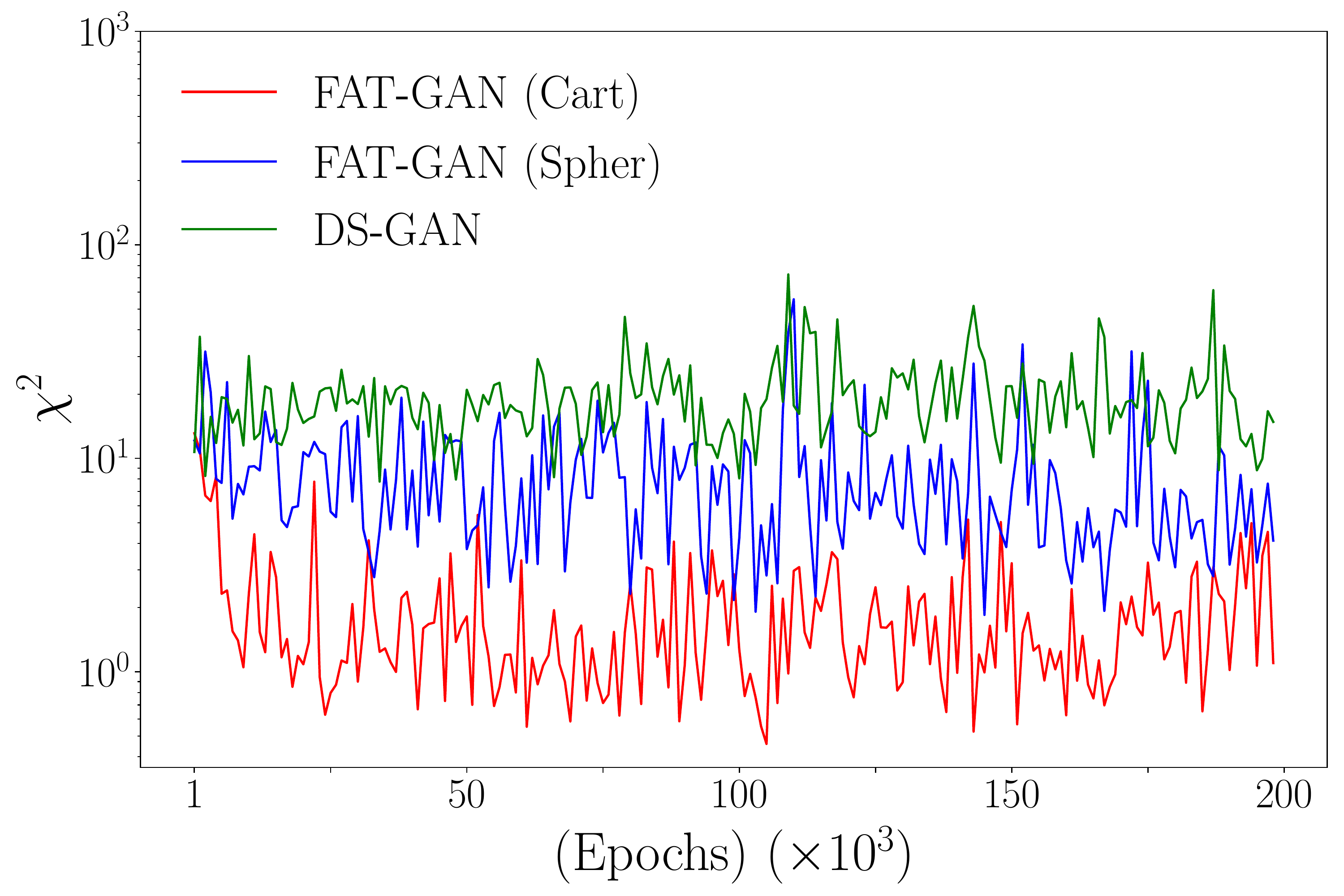}
  \captionof{figure}{Comparison of $\chi^2$ values for $x_{\rm bj}$ distributions of Pythia events and those generated by the FAT-GAN (Cart), FAT-GAN (Spher) and DS-GAN, with respect to the number of training epochs.}
  \label{fig:chi2}
\end{center}

Overall, the FAT-GAN (Spher) is found to have a degraded performance compared to the FAT-GAN (Cart).
The main reason is that the physical properties in spherical coordinates are less favorable than those in Cartesian coordinates for training the FAT-GAN.
In particular, both the distributions of $E$ and $\theta$ in spherical coordinates exhibit sharp edges (see Fig.~\ref{fig:GANresults}(d) and (f)).
Moreover, the $\phi$ distribution has shape boundaries at both ends, which poses additional complications for the GAN to learn.

\subsection{FAT-GAN on experimental data}
\label{section:Experiment}

Having studied the performance of the FAT-GAN on synthetic Pythia data, we now compare the results of the FAT-GAN (Cart) simulations with inclusive scattering data for a 5.5~GeV electron beam impinging on a liquid hydrogen target in CLAS at Jefferson Lab~\cite{Mecking:2003zu}.
A clean event sample of 100k reconstructed electron events was used for the comparison with the trained FAT-GAN, illustrated in Fig.~\ref{fig:exp}, which shows that the distributions of the events generated by FAT-GAN match well with those of the experimental data.
Note, however, that in contrast to the smooth distributions from the Pythia MCEG in Fig.~\ref{fig:GANresults}, the CLAS data here have not been corrected for detector effects, and therefore display artifacts such as the shoulder in the $x_{\rm bj}$ distribution.
Nevertheless, the comparison in Fig.~\ref{fig:exp} indicates that the physical properties $p_T$, $\theta$, $x_{\rm bj}$, and $Q^2$, which are not directly used in the training, are correctly captured by the FAT-GAN trained on the experimental event data.

\begin{center}
  \includegraphics[width=\linewidth]{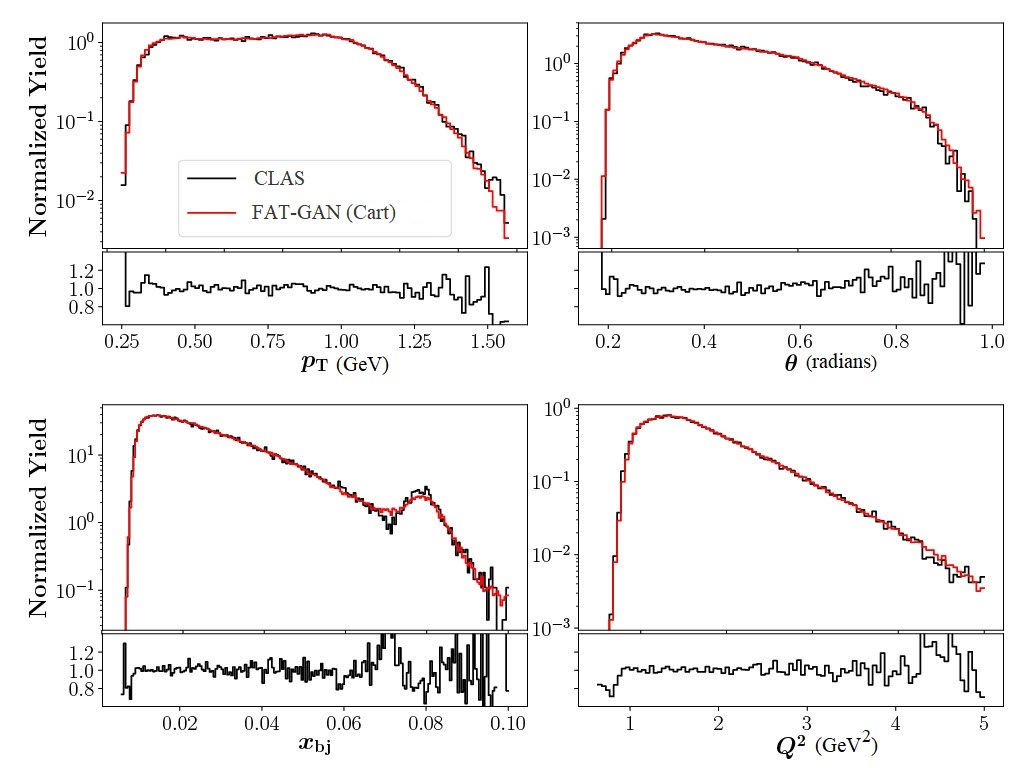}
  \captionof{figure}{Comparison of the distributions of the physical properties $p_T$, $\theta$, $x_{\rm bj}$ and $Q^2$ generated by FAT-GAN (Cart) with those from CLAS at Jefferson Lab (uncorrected for detector effects). The ratio of FAT-GAN (Cart) to CLAS data is shown at the bottom of each panel.}
  \label{fig:exp}
\end{center}

\section{Conclusion}

We have investigated the use of GANs to simulate electrons in the final state of high-energy inclusive electron-proton collisions.
We report that selecting the appropriate features as generated features or augmented features plays a critical role in building a successful GAN event generator.
While the physical properties of the particles often exhibit challenging distribution patterns for a GAN to learn, using transformed features enables the generator to more easily generate and avoid unphysical events.
Augmenting in addition the feature space for the discriminator to become more sensitive, our FAT-GAN demonstrates good agreement with simulated as well as experimental data in mimicking event feature distributions and their inter-correlations.
We also find that the selection of the appropriate coordinate representations impacts the GAN performance. 
Although the FAT-GANs presented in this paper are specific to electron-proton scattering, the feature selections and transformation strategy can be generalized to GANs for simulating other reactions under different conditions, as well as learning exclusive events. 

The FAT-GAN package is available at \url{https://github.com/ijcai2021/FAT-GAN}.

\section*{Acknowledgements}
We thank the CLAS Collaboration for providing the electron scattering data used in Fig.~4, and J.~Qiu for helpful discussions.
This work was supported by the LDRD project No.~LDRD19-13, No.~LDRD20-18, and No.~LDRD21-22.

\appendix

\bibliographystyle{named}
\bibliography{ijcai21}
\end{document}